\documentclass[aps,prd,twocolumn,amsmath,groupedaddress,nofootinbib]{revtex4-1}
\usepackage{amsmath}
\usepackage{amssymb}
\usepackage{latexsym}
\usepackage{graphicx}
\usepackage{hyperref}
\usepackage{mathrsfs}
\usepackage{color}
\usepackage{bm}
\usepackage[caption=false]{subfig}

\def\bhD{\mathscr{D}_{\bullet}}
\def\bhm{M_{\bullet}}

\def\mathdotM{\dot{\mathscr{M}}}
\def\ergs{erg\,s$^{-1}$}
\def\sunm{M_{\odot}}

\begin{document}

\title{Super-Eddington accreting massive black holes explore high-$z$ cosmology: Monte-Carlo simulations}

\author{Rong-Gen Cai}
\email{cairg@itp.ac.cn}

\author{Zong-Kuan Guo}
\email{guozk@itp.ac.cn}

\author{Qing-Guo Huang}
\email{huangqg@itp.ac.cn}

\affiliation{CAS Key Laboratory of Theoretical Physics, Institute of Theoretical Physics, Chinese Academy of Sciences, Beijing 100190, China}
\affiliation{School of Physical Sciences, University of Chinese Academy of Sciences, No.19A Yuquan Road, Beijing 100049, China}

\author{Tao Yang}
\email{yangtao2017@bnu.edu.cn}

\affiliation{Department of Astronomy, Beijing Normal University, Beijing, 100875, China}

\date{\today}

\begin{abstract}
In this paper, we simulate Super-Eddington accreting massive black holes (SEAMBHs) as the
candles to probe cosmology for the first time. SEAMBHs have been demonstrated to be able to
provide a new tool for estimating cosmological distance. Thus, we create a series of mock data
sets of SEAMBHs, especially in the high redshift region, to check their abilities to probe
the cosmology. To fulfill the potential of the SEAMBHs on the cosmology, we apply the simulated
data to three projects. The first is the exploration of their abilities to constrain the cosmological
parameters, in which we combine different data sets of current observations such as the cosmic
microwave background from {\it Planck} and type Ia supernovae from Joint
Light-curve Analysis (JLA). We find that the high redshift SEAMBHs can help to break the
degeneracies of the background cosmological parameters constrained by {\it Planck} and JLA,
thus giving much tighter constraints of the cosmological parameters. The second uses the
high redshift SEAMBHs as the complements of the low redshift JLA to constrain the early
expansion rate and the dark energy density evolution in the cold dark matter frame. Our results show that these high redshift SEAMBHs are very powerful on
constraining the early Hubble rate and the evolution of the dark energy density; thus they can give us more information about the expansion
history of our Universe, which is also crucial for testing the $\Lambda$CDM model in the high
redshift region. Finally, we  check the SEAMBH candles' abilities to reconstruct the
equation of state of dark energy at high redshift. In summary, our results show that the
SEAMBHs, as the rare candles in the high redshift region, can provide us a new and independent
observation to probe cosmology in the future.

\end{abstract}

\maketitle

\section{Introduction}
\label{sec:int}

The accelerating expansion of the Universe has been discovered from the studies of distant
Type Ia supernovae (SNe Ia) by two teams~\cite{Riess:1998cb,Perlmutter:1998np} for nearly
twenty years. These two decades have also witnessed  rapid technological advances
in observational cosmology. Various observations such as the
SNe Ia~\cite{Suzuki:2011hu,Betoule:2014frx}, the temperature and polarization anisotropy
power spectrum of the cosmic microwave background (CMB) radiation~\cite{Hinshaw:2012aka,Ade:2015xua},
 and baryon acoustic oscillations (BAO)~\cite{Beutler:2011hx,Ross:2014qpa,Anderson:2013zyy}
have all suggested that the late time Universe is dominated by a simple dark energy component, for which a simple
explanation  is the cosmological constant $\Lambda$. The tiny cosmological constant with a constant equation of
state $-1$ combined with cold dark matte (called $\Lambda$CDM) turns out to be the standard
model which fits the current observational data sets consistently.

However, the $\Lambda$CDM model is faced with the fine-tuning problem and the coincidence
problem~\cite{Weinberg:2000yb}. The former arises from the fact that the present-time observed
value for the vacuum energy density is more than 120 orders of magnitude smaller than the naive
estimate from quantum field theory. The latter is the question why we live in such a special
moment that the densities of dark energy and dark matter are of the same order. Many attempts
have been made to tackle these issues, including introducing ``dynamical" dark energy.
Moreover, some of the different data sets are not so well consistent among them. For example, there is a
strong tension between the value of the Hubble constant derived from the CMB~\cite{Ade:2015xua}
and the value from local measurements~\cite{Riess:2011yx}. Thus, understanding the physical
properties of dark energy, such as whether it is dynamical ($w\neq-1$) or not, is one of the
main challenges of modern cosmology.

SNe Ia as the standard candles are powerful probes of cosmology and in particular to the
equation of state of dark energy.
However, current data sets of SNe Ia such as the ``Union 2.1" compilation~\cite{Suzuki:2011hu}
and the ``Joint Light-curve Analysis" (JLA) sample~\cite{Betoule:2014frx} are mainly concentrated
in the low redshift ($z<1$) region. Though the constant $w$ can be constrained well enough using
$z<1$ SNe Ia, if we want to study the dynamics of dark energy such as the time-varying $w$,
or  to constrain the expansion rate of the Universe and check the $\Lambda$CDM model back to earlier time,
the high redshift observations thus become crucially necessary.
For example, recently, Riess {\it et al.}~\cite{Riess:2017lxs} presented an analysis of 15 Type Ia
supernovae at redshift $z>1$ (9 at $1.5 < z < 2.3$) discovered in the CANDELS and CLASH
Multi-Cycle Treasury programs using WFC3 on the {\it Hubble Space Telescope}. They found that the
added leverage of these new samples at $z > 1.5$ leads to a factor of $\sim 3$ improvement in the
determination of the expansion rate at $z = 1.5$, reducing its uncertainty to $\sim20\%$.
High-$z$ SNe Ia are rare since the capability of instruments and intrinsic evolution of progenitors
of SNe Ia in future surveys as shown by Hook {\it et al.}~\cite{Hook:2012xk}.

Recently, a new kind of cosmic long-lived candles employing super-Eddington accreting massive black holes (SEMABHs)
has been suggested for measurements of expansion rates of the Universe
by Wang {\it et al.} since they are characterized by the saturated luminosity
(see~\cite{Wang:2013ha,Wang:2014fka} and references therein).
The advantages of this new tool
are (see Sec.~\ref{sec:SEAMBHs} for details):
1) their abundance increases with redshifts and about $10\%-30\%$ of quasars in the local Universe contain SEAMBHs~\cite{Du:2016ApJL,Du:2016egf};
2) the principle of the SEAMBHs relies on the saturated luminosity, which results from the well understood photon trapping effects
in super-Eddington accretion onto black holes~\cite{Abramowicz:1988sp,Wang:1999apj516,Mineshige:2000tf}. This is
well-understood in theory and is observationally tested
by a long-term reverberation mapping campaign of spectroscopically monitoring SEAMBHs~\cite{Du:2013kya,Du:2015yka,Du:2016vrl}.
SEAMBHs can be considered as candles similar to SNe Ia.
While the advantage of SEAMBHs here is that they can extend the SN-based information of the expansion history of the universe to a much higher redshift, $z\sim 6$, than previously possible.  Thus, it is possible for us to use these high redshift SEAMBHs to probe the cosmology. In this paper, we simulate the SEAMBHs distance-redshift data sets at high redshifts from $z\sim 1$ to 6. We use these mock data sets to forecast the abilities of future SEAMBHs candles to probe the cosmology in three different schemes. The first is using these SEAMBHs combined with different data sets such as the {\it Planck} and JLA to constrain the cosmological parameters in the Chevallier-Polarski-Linder (CPL) parametric dark energy model.  We want to demonstrate the roles of these high redshift SEAMBHs  in constraining the background cosmological parameters and to show how many improvements they can provide to the {\it Planck} + JLA combination. The second is using the high redshift SEAMBHs as the complements of the low redshift JLA to constrain the early expansion rate and to compare it with JLA. Finally, we  try to reconstruct the equation of state using these SEAMBHs and JLA data sets.

The paper is organized as follows. In Sec.~\ref{sec:SEAMBHs} we give brief introductions of the SEAMBHs and their properties as new cosmological standards. In Sec.~\ref{sec:simulations}, the strategy of the simulation of the SEAMBHs data sets will be outlined. Then, in Sec.~\ref{sec:P1} to~\ref{sec:P3} we will apply the mock data sets of SEAMBHs to probing the cosmology in three schemes and all of the results will be shown. Finally, the conclusions and discussions will be given in Sec.~\ref{sec:conclusions}.

\section{Super-Eddington accreting massive black holes as candles}
\label{sec:SEAMBHs}

Giant gravitational energy is released by accretion onto black holes (BHs)~\cite{Shakura:1972te,Pringle:1981ds} and
powers luminous active galactic nuclei and quasars~\cite{Rees:1984si}. It is well understood that the
radiative luminosity from accretion disks is linearly proportional to accretion rates, known as the so-called standard
accretion disks with rates of $\dot{M}\lesssim 3.0\dot{M}_{\rm Edd}$~\cite{Pringle:1981ds},  and the disks keep Keplerian
rotation around the black hole, and the radial motion can be neglected compared with the Keplerian velocity.
Here $\dot{M}_{\rm Edd}=L_{\rm Edd}/c^2$,
where $L_{\rm Edd}=1.5\times 10^{38}\left(\bhm/\sunm\right)$\ergs\, is the Eddington luminosity, $c$ is the  speed of light
and $\bhm$ is the BH mass (see Figure 1 in~\cite{Abramowicz:1988sp}).

However, the radial motion of accretion
flows becomes very important when $\dot{M}\gtrsim3.0\dot{M}_{\rm Edd}$~\cite{Laor:1989}
and the energy balance is globalized so that most of photons
produced from the viscous dissipation are advocated into black holes before they escape from the disk surface. This is caused by
a very large Thompson scattering depth in vertical direction when $\dot{M}\gtrsim 3.0\dot{M}_{\rm Edd}$. This is the basic idea of slim accretion disks~\cite{Abramowicz:1988sp}. The photon trapping effect gives rise to the most prominent feature known
as the saturated luminosity, which is only linearly proportional to
the black hole mass and very insensitive to accretion rates. The self-similar solution of extreme slim disks shows
\begin{equation}
L_{\bullet}=\ell_0(1+a\ln\mathdotM)L_{\rm Edd}\,,
\label{equa:L}
\end{equation}
where $\ell_0\approx (2-4)$ and $a \lesssim 0.5$, both of which depend on the vertical structure of the slim disks~\cite{Wang:1999apj516,Mineshige:2000tf}. It was a purely theoretical result of slim disks, but the long-term reverberation mapping
campaigns of SEAMBH project lend unambiguous support to the saturated luminosity ~\cite{Du:2015yka,Du:2016egf}. Though Eq.~(\ref{equa:L}) shows a logarithmic dependence on accretion rates, an observational test shows the
saturated luminosity is much weaker than the logarithmic~\cite{Du:2015yka,Du:2016egf}.

The scheme of determining cosmological distance has been outlined by Wang {\it et al.}~\cite{Wang:2014fka}.
The distance is given by
\begin{equation}
\mathscr{D}_{\bullet}=250.3\,\ell_{\kappa}^{1/2}m_7^{(1+\beta)/2}F_{11}^{-1/2}\,{\rm Mpc}\,,
\label{equa:mathD}
\end{equation}
where $\ell_{\kappa}$ is the factor including the bolometric correction factor, inclination of accretion disks
$\cos i \approx 0.75$, and black hole spins,
$\beta\approx0.3\sim0.4$ is the exponential index of the
dependence of bolometric correction
factor on BH mass,
$m_7=\bhm/10^7\sunm$ and $F_{11}=F_{5100}/10^{-11}\,{\rm erg\,s^{-1}\,cm^{-2}}$.
The accuracy of the distance measured by the black hole candles is mainly determined by the accuracy of black hole mass since
the factor $\ell_{\kappa}$ can be calibrated by the local distance. According to Eq.~(\ref{equa:mathD}), we have
\begin{equation}
\frac{\Delta \bhD}{\bhD}=\left(\frac{1+\beta}{2}\right)\frac{\Delta \bhm}{\bhm}\,.
\end{equation}
The key of using SEAMBH as powerful indicator of cosmological distances depends on the accuracy of black hole mass.

\def\rblr{R_{\rm BLR}}
\def\fblr{f_{\rm BLR}}

Currently, the black hole mass is estimated by the virial relation of $\bhm=\fblr G^{-1}R_{\rm BLR}V_{\rm FWHM}^2$,
where $\fblr$ is the virial factor, $G$ is gravitational constant,
$R_{\rm BLR}$ is the reverberation radius of the the broad-line regions and
$V_{\rm FWHM}$ is the full-width-half-maximum of the H$\beta$ profiles. Three approaches to estimation of black hole
mass reach different accuracies as briefly discussed below. The accuracy can be estimated by
\begin{equation}
\frac{\Delta \bhm}{\bhm}\approx \sqrt{\left(\frac{\Delta \fblr}{\fblr}\right)^2
                                      +\left(\frac{\Delta \rblr}{\rblr}\right)^2
                                      +2\left(\frac{\Delta V_{\rm FWHM}}{V_{\rm FWHM}}\right)^2}\,,
\end{equation}
where $\Delta\fblr/\fblr\sim 0.5$, due to inclinations and geometries~\cite{Ho:2014pka,Mejia-Restrepo:2017pqs}.
The major uncertainties are from the fact that we do not know if the observed $\rblr$ regions are the virialized
component (i.e. if it corresponds to $V_{\rm FWHM}$). Direct measurements using reverberation mapping of AGNs yield
an accuracy of $\Delta\bhm/\bhm\sim 1$. Using empirical
relation between $R_{\rm BLR}$ and optical luminosity~\cite{Kaspi:1999pz,Bentz:2013wxa,Du:2015yka,Du:2016egf}, we
will have additional error bars from the scatters of the relation as much as $0.2$ dex~\cite{Bentz:2013wxa}, but
will be larger for SEAMBHs~\cite{Du:2015yka,Du:2016egf}. It has been recently suggested by Wang {\it et al.}~\cite{Wang:2017Natas}
that the total profiles of H$\beta$ emission line can be physically separated to find the virialized components. This
can greatly reduce uncertainties of $\Delta \rblr$ and $\Delta V_{\rm FWHM}$ and improve the accuracy of
the black hole mass. More accurately,
a Markov Chain Monte Carlo  (MCMC) is employed to model the light curves and profiles simultaneously~\cite{Pancoast:2012pm,Li:2013qua,Grier:2017} and reaches an accuracy of 10\% for some individuals.

On the other hand,
$f_{\rm BLR}$ can be well constrained by error bars of a few percent
from the polarized spectra~\cite{Songsheng:2018}. Precision estimation of BH mass remains open, but it can
reach 50\%~\footnote{Considering the dusty torus, type I AGNs have inclinations of $i\le 45^{\circ}$, otherwise they appear as type II. In fact the uncertainties of inclinations are quiet small.  The accuracy of 50\% covers the uncertainties of inclinations.} or so as a conservative value on average for a large sample, and better than 10\% from the polarized
spectra of individual SEAMBHs. In this paper, we presume the error bar of $50\%$ for the large amount of the
SEAMBH candles in the hight redshift region ($z\sim1-6$).
While we also consider a small amount (50 or 100) of SEAMBHs
candles at redshift $z\in[1,2]$ as the high precise measurements through direct reverberation mapping campaigns,
we  employ them to establish better empirical relations for black hole masses.

We would point out that, contrary to SNe Ia, SEAMBH numbers increase with redshifts, in particular,
they have a large fraction increasing with redshifts from data of the Sloan Digital Sky Survey (see their Figure 4 in~\cite{Kelly:2012vz} by Kelly \& Shen). It should be noted that the Kelly \& Shen's
estimations of SEAMBH numbers are conservative since they use the canonical $R-L$ relation overestimating the black
hole mass for SEAMBHs~\cite{Du:2016ApJL}. Future spectroscopic survey of Dark Energy Survey
Instrument~\cite{Aghamousa:2016zmz} will find much more SEAMBHs than the current
SDSS. The presumed SEAMBH number cross redshifts is actually feasible.

Large scale campaigns of reverberation mapping of AGNs and quasars have been started from SDSS ~\cite{Shen:2014uby} and obtained preliminary results~\cite{Grier:2017xel}. The SDSS campaigns employ a spectrography with 600 fibers so that it is much more efficient than the traditional ones. It is more exciting in the near future that Maunakea Spectroscopic Explorer
(MSE~\footnote{\url{http://mse.cfht.hawaii.edu/project/}}) as a 10m telescope with 3000 fibers-fed spectrography will do reverberation mapping campaigns of AGNs. This greatly
increases the numbers and mass accuracy of SEAMBHs, and hence the feasibility of SEAMBHs for cosmology.

\section{The simulations of the SEAMBH distance-redshift data sets}
\label{sec:simulations}

From the properties of the SEAMBHs described in Sec.~\ref{sec:SEAMBHs}, we outline our strategies of simulating the SEAMBHs distance-redshift data sets here. For a  Friedmann-Robertson-Walker (FRW) universe, the luminosity distance can be written as
\begin{equation}
{d_L} =\begin{cases}
\frac{{(1 + z)}}{{{H_0}\sqrt {\Omega_K}}}\sinh (\sqrt {\Omega_K} \int_0^z {\frac{{d\tilde z}}{{E(\tilde z)}})} & \Omega_K > 0 \\
\frac{{(1 + z)}}{{H_0}}\int_0^z {\frac{{d\tilde z}}{{E(\tilde z)}}} & \Omega_K = 0 \\
\frac{{(1 + z)}}{{{H_0}\sqrt {\left|\Omega_K\right|}}}\sin (\sqrt {\left|\Omega_K\right|} \int_0^z {\frac{{d\tilde z}}{{E(\tilde z)}})} & \Omega_K < 0\,,
\end{cases}
\label{equa:dl}
\end{equation}
where $E(z)\equiv H(z)/H_0$ is the Hubble rate, and $\Omega_K\equiv -K/(a_0 H_0)^2$ is the cosmic curvature today. With the  dark energy equation of state $w(z)=p(z)/\rho(z)$, the Hubble parameter $H(z)$ is given by Friedmann equation,
\begin{align}
H{(z)^2} = &~~H_0^2 \Big\{ (1 - {\Omega _m} - {\Omega _K})\exp \Big[3\int_0^z {\frac{{1 + w(\tilde z)}}{{1 + \tilde z}}} d\tilde z\Big] \nonumber\\
           &~~+{\Omega _m}{(1 + z)^3} + {\Omega _K}{(1 + z)^2}\Big\}\,,
\label{equa:H}
\end{align}
where $\Omega_{m,K}$ are the matter and curvature density parameters today.
Combining Eqs.~(\ref{equa:dl}) and~(\ref{equa:H}) and writing $D(z)=H_0(1+z)^{-1} d_L(z)$ as the normalized comoving distance, we find that the equation of state can be expressed as
\begin{align}
w(z) = &~~\Big[2(1 + z)(1 + {\Omega _K}{D^2})D'' - {(1 + z)^2}{\Omega _K}D{'^3} \nonumber\\
       &~~-2(1+z){\Omega_K}DD{'^2}+3(1+{\Omega_K}D^2)D'\Big] \nonumber\\
       &~~\Big/\Big[3\{ {(1 + z)^2}[{\Omega _K} + (1 + z){\Omega _m}]D{'^2} \nonumber\\
       &~~- (1 + {\Omega _K}{D^2})\} D'\Big]\,.
\label{equa:w}
\end{align}

Given the high quality of the SNe Ia data such as JLA in the low redshift region (from $z \sim 0-1$) and the advantages of our SEAMBHs candles at high redshifts, we mainly focus on forecasting the abilities and improvements of the SEAMBHs candles to probe the cosmology at redshifts lager than 1.
From the estimation of the practical observations in Sec.~\ref{sec:SEAMBHs}, we simulate the distributions of the SEAMBHs redshifts and their corresponding uncertainties of distance measurements according to different research object and numerical technique. We divide the mock data sets generating procedure into two parts:

\textbf{Part I}: These mock data sets are used in the first scheme, that is adopting the Markov Chain Monte Carlo and combining these mock data sets with {\it Planck} + JLA to constrain the cosmological parameters.

\begin{itemize}
	\item {\bf Case I}: Redshift $z \sim (1,2)$; Number $\sim 50$ (or 100); Precision $\sim20\%$;
	\item {\bf Case II}: Redshift $z \sim (1,2)$; Number $\sim 5000$ (or 10000); Precision $\sim50\%$;
	\item {\bf Case III}: Redshift $z \sim (2,4)$; Number $\sim 5000$ (or 10000); Precision $\sim50\%$;
    \item {\bf Case IV}: Redshift $z \sim (4,6)$; Number $\sim 1000$; Precision $\sim50\%$;
\end{itemize}

\textbf{Part II}: These mock data sets are used in the second and third projects of this work. Since the number of data points is limited by the numerical technique Gaussian Process which is used in these two project, we reduce the number of the data points for the cases of mock data. We also give a binned data sets which represent the SEAMBHs' most complete data sets based on current estimation.

\begin{itemize}
	\item {\bf Case I}: Redshift $z \sim (1,2)$; Number $\sim 50$; Precision $\sim20\%$;
	\item {\bf Case II}: Redshift $z \sim (1,2)$; Number $\sim 1000$; Precision $\sim50\%$;
	\item {\bf Case III}: Redshift $z \sim (2,4)$; Number $\sim 1000$; Precision $\sim50\%$;
    \item {\bf Case all bin}: The most complete data sets based on current estimation. The redshift is from 1 to 6, the total number of points is \{$z\sim(1,2):50\} + \{z\sim(1,2):10000\} + \{z\sim(2,3):10000\} + \{z\sim(3,4):5000\} + \{z\sim(4,5):1000\} + \{z\sim(5,6):500$\}. The uncertainties are the same strategy as before. However, we bin these data and improve the uncertainties according to $\sqrt{N}$ and then we use the binned data instead.
\end{itemize}

Note that since we just simulate the future measurements, as a rough estimation, the precision here covers all uncertainties from the calibration and systematic errors and so on.
For the simulations of the data sets, we have to choose a fiducial cosmological model. The exact values of the cosmological parameters will not be essential in our simulations, because we are just interested in the precision with which they can be measured. However, for consistency with the current experiment data {\it Planck} 2015~\cite{Ade:2015xua}, we choose the cosmological parameters of the fiducial model as follows,
\begin{align}
h_0=0.678,~~\Omega_m=0.308,~~\Omega_K=0,~~w=-1\,,
\label{ini}
\end{align}
here $H_0=100h_0$km\,s$^{-1}$Mpc$^{-1}$. When we combine the SEAMBHs mock data with only the JLA data sets in constraining the Hubble rate and dark energy density and also reconstructing the equation of state, we choose a slightly different fiducial value of $h_0=0.7$ and $\Omega_m=0.3$ to be more
compatible with JLA. Anyway, the slightly different choices of the fiducial value will not influence our results.

\section{Project I: constrain the cosmological parameters combined with {\it Planck} and JLA}
\label{sec:P1}

In this section, we use the high redshift mock data sets of BH candles (hereafter we also refer to SEAMBH candles briefly as BH candles) combined with low redshift JLA data and the {\it Planck} data ($Planck~TT$ + lowP) to see how much improvement these BH candles can provide to the constraints of the cosmological parameters by the SNIa standard candles and {\it Planck}.

\begin{figure*}
\includegraphics[width=0.45\textwidth]{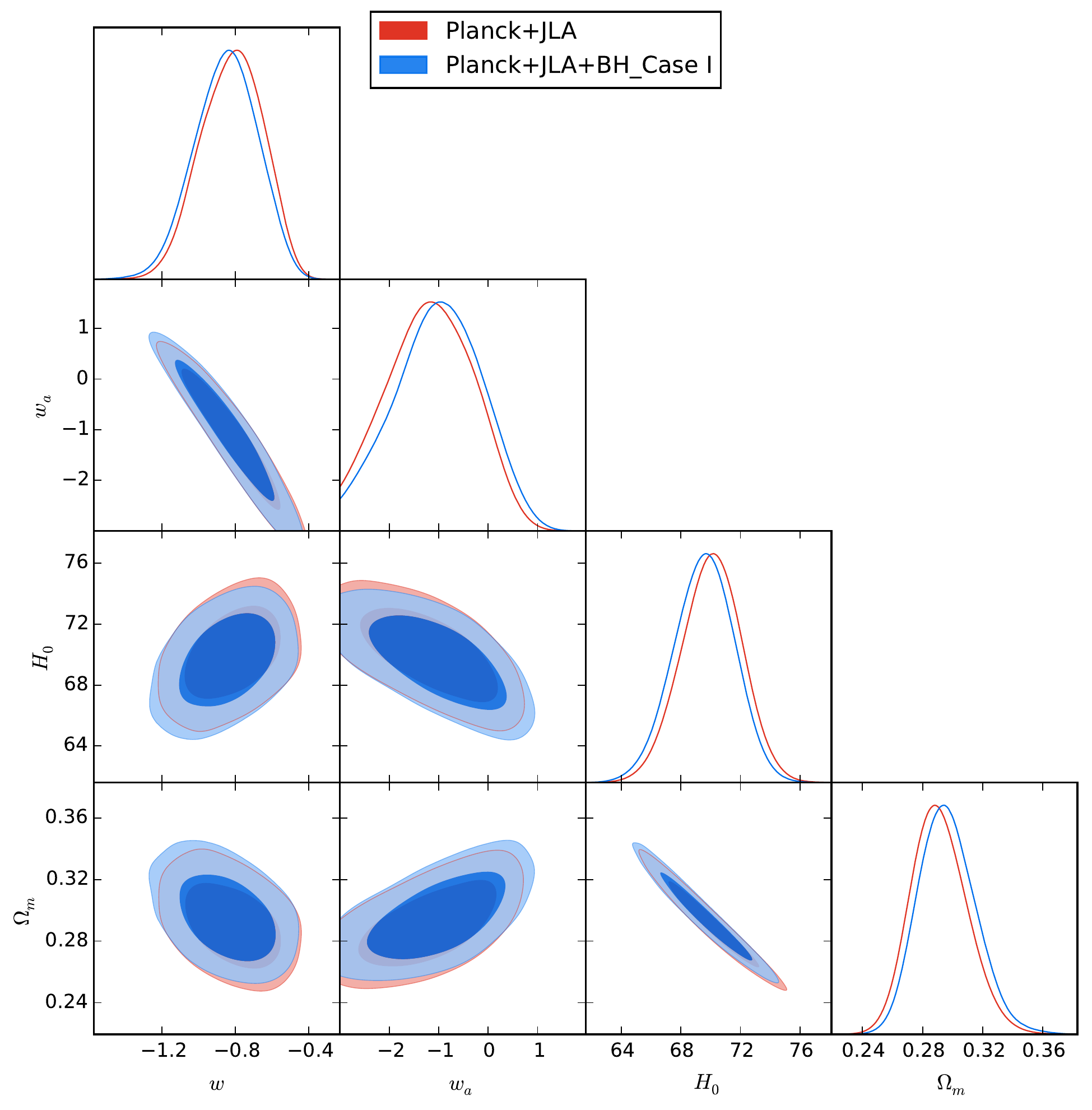}
\includegraphics[width=0.45\textwidth]{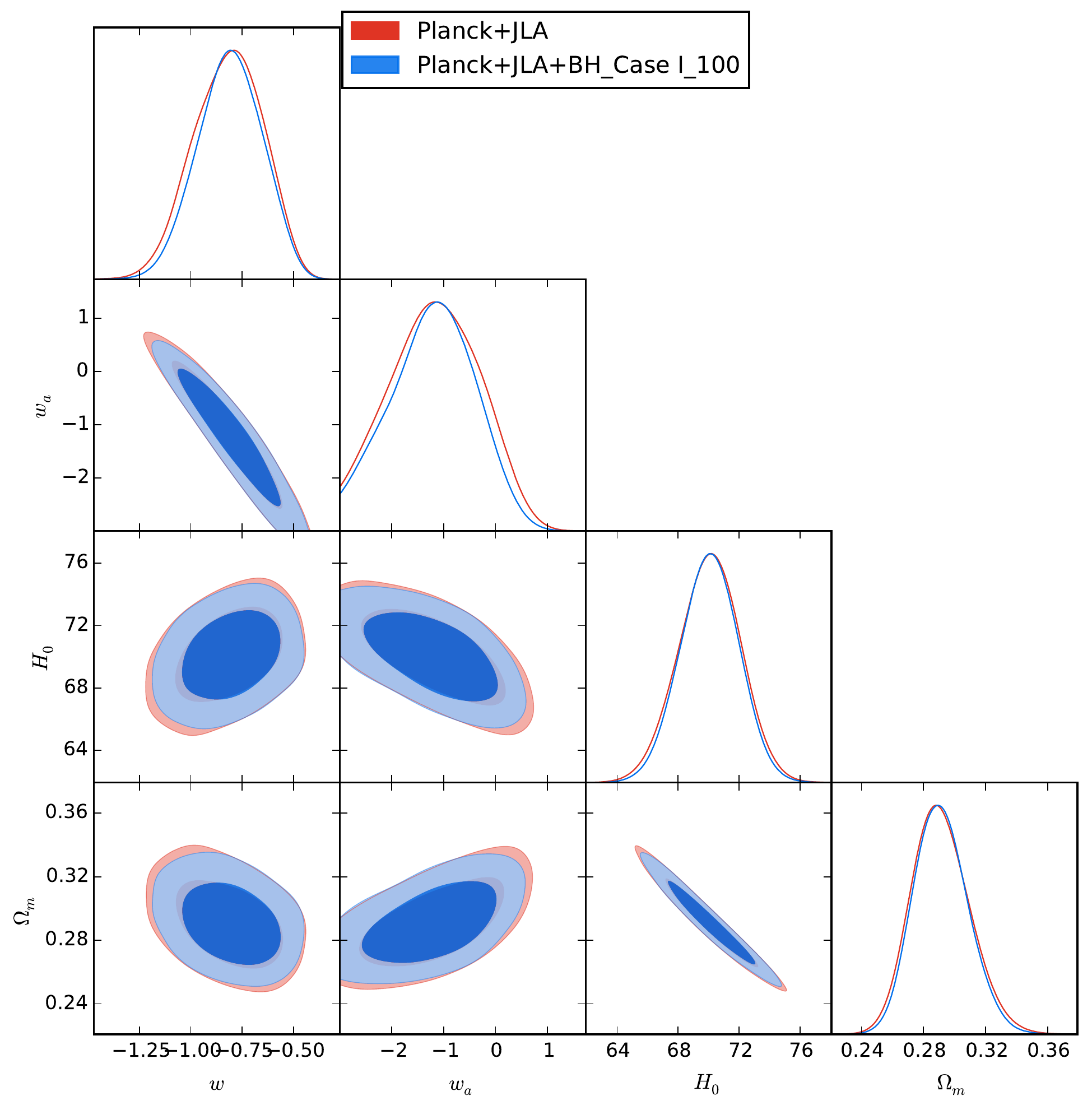}\\
\caption{The comparison of the constraints of the cosmological parameters from different data set combinations. The BH mock data sets are the case I (left) and case I but with increasing the number of data points from 50 to 100 (right).}
\label{fig:PJB_I}
\end{figure*}

\begin{figure*}
\includegraphics[width=0.45\textwidth]{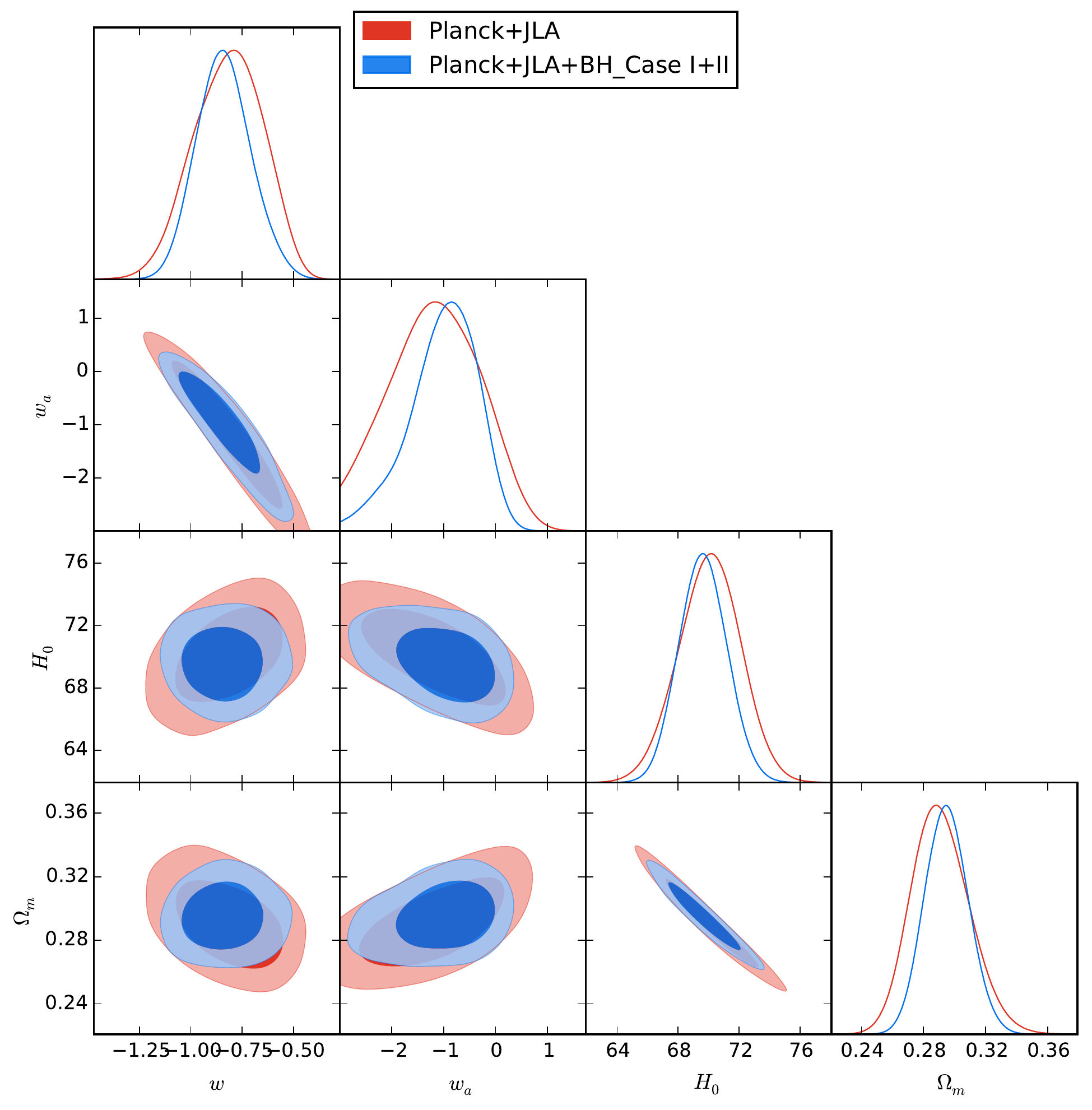}
\includegraphics[width=0.45\textwidth]{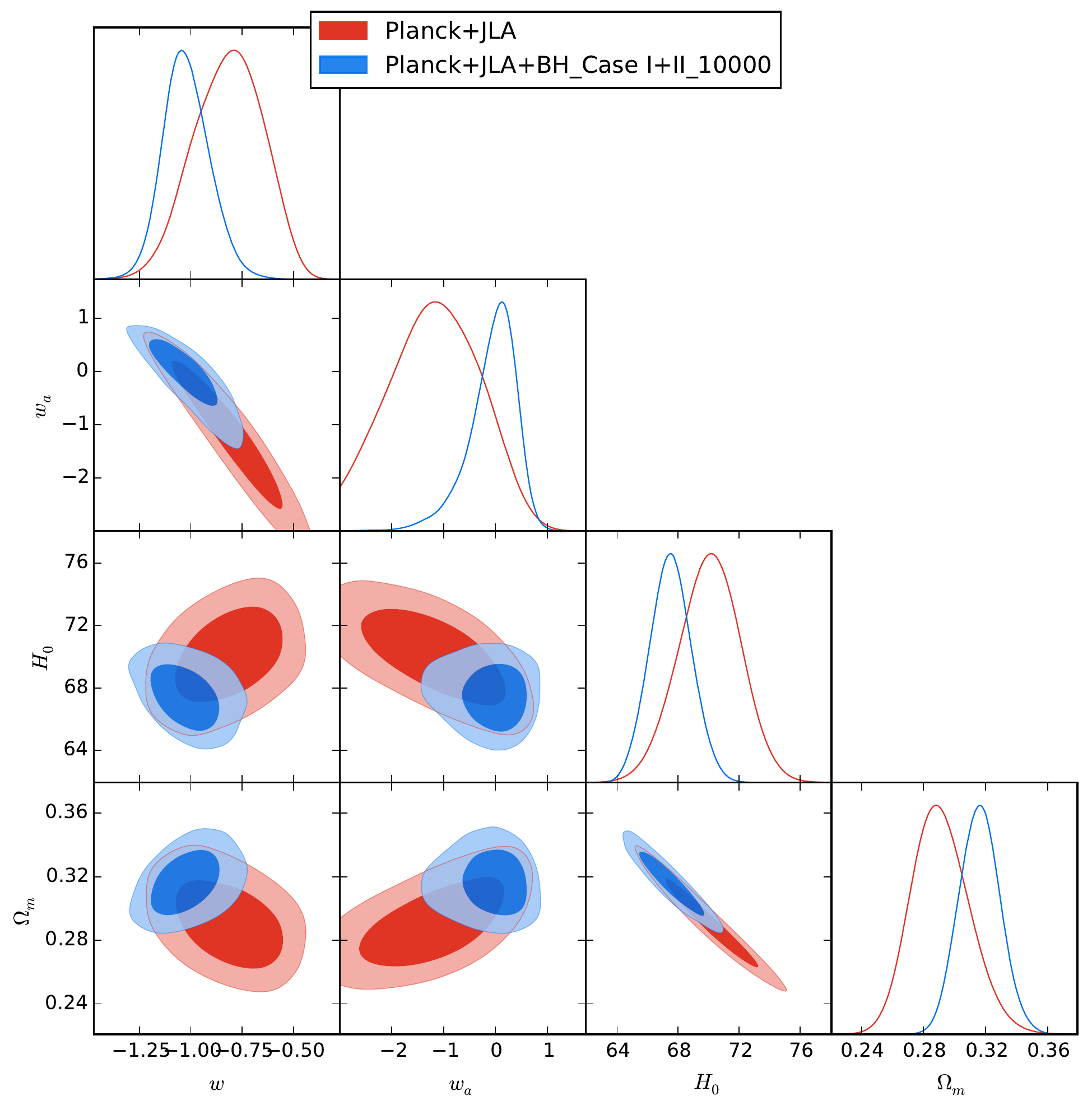}
\caption{The comparison of the constraints of the cosmological parameters from different data set combinations. The BH mock data sets are the case I + II (left) and case I + II but increasing the number of data points of case II from 5000 to 10000 (right).}
\label{fig:PJB_II}
\end{figure*}

\begin{figure*}
\includegraphics[width=0.45\textwidth]{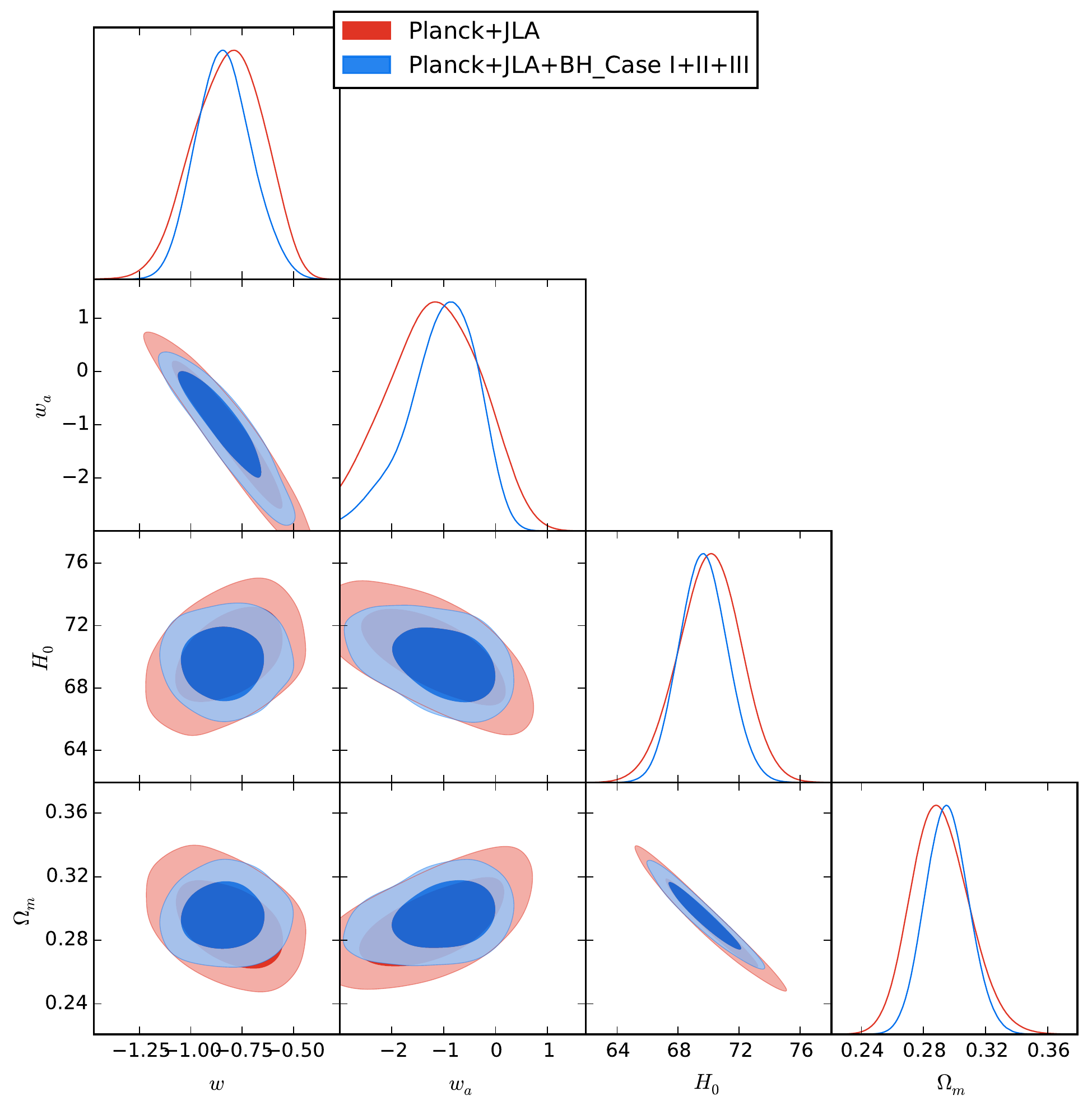}
\includegraphics[width=0.45\textwidth]{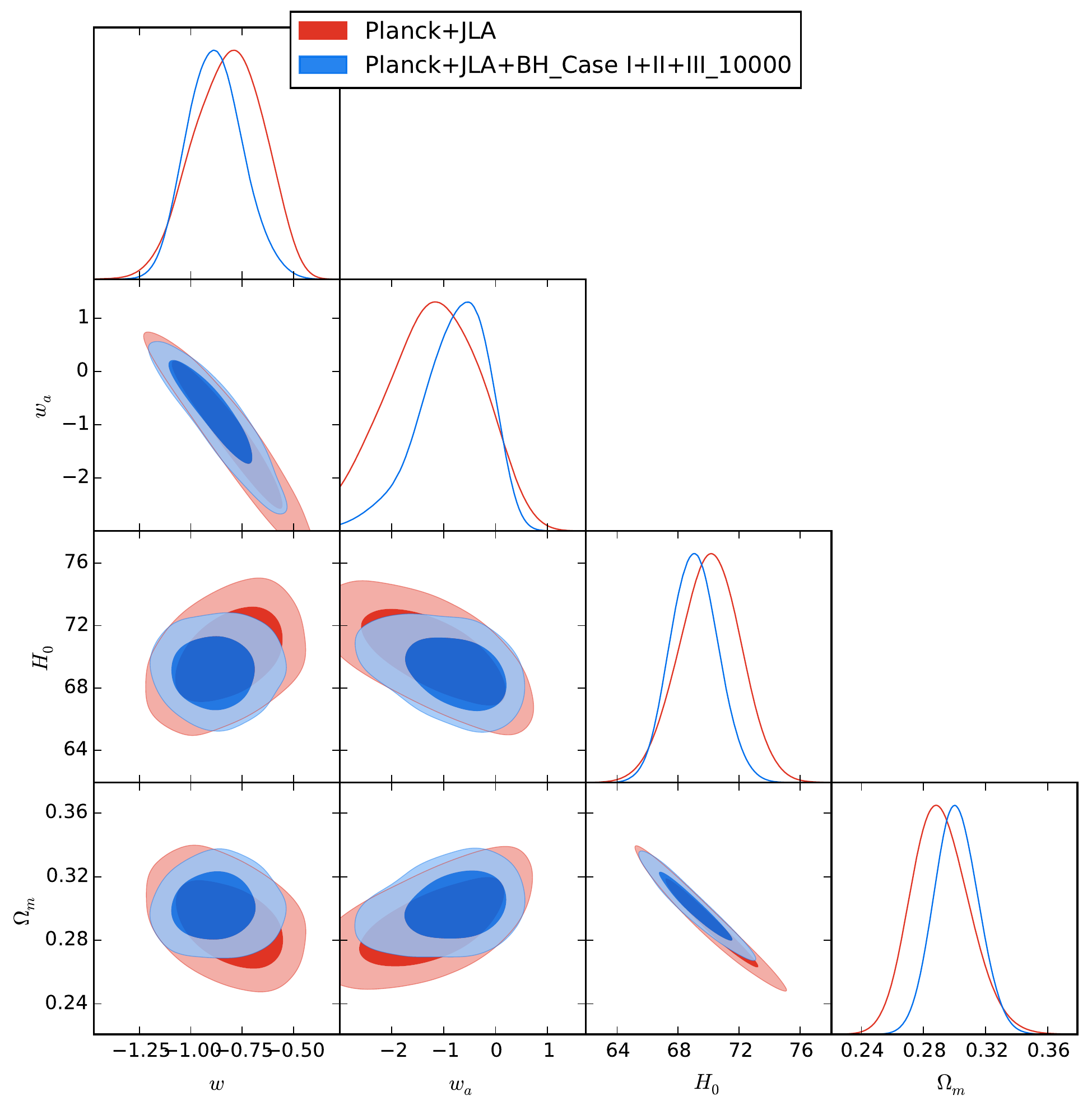}
\caption{The comparison of the constraints of the cosmological parameters from different data sets combinations. The BH mock data sets are the case I + II + III (left) and case I + II + III but increasing the number of data points of case III from 5000 to 10000 (right).}
\label{fig:PJB_III}
\end{figure*}

\begin{figure*}
\includegraphics[width=0.45\textwidth]{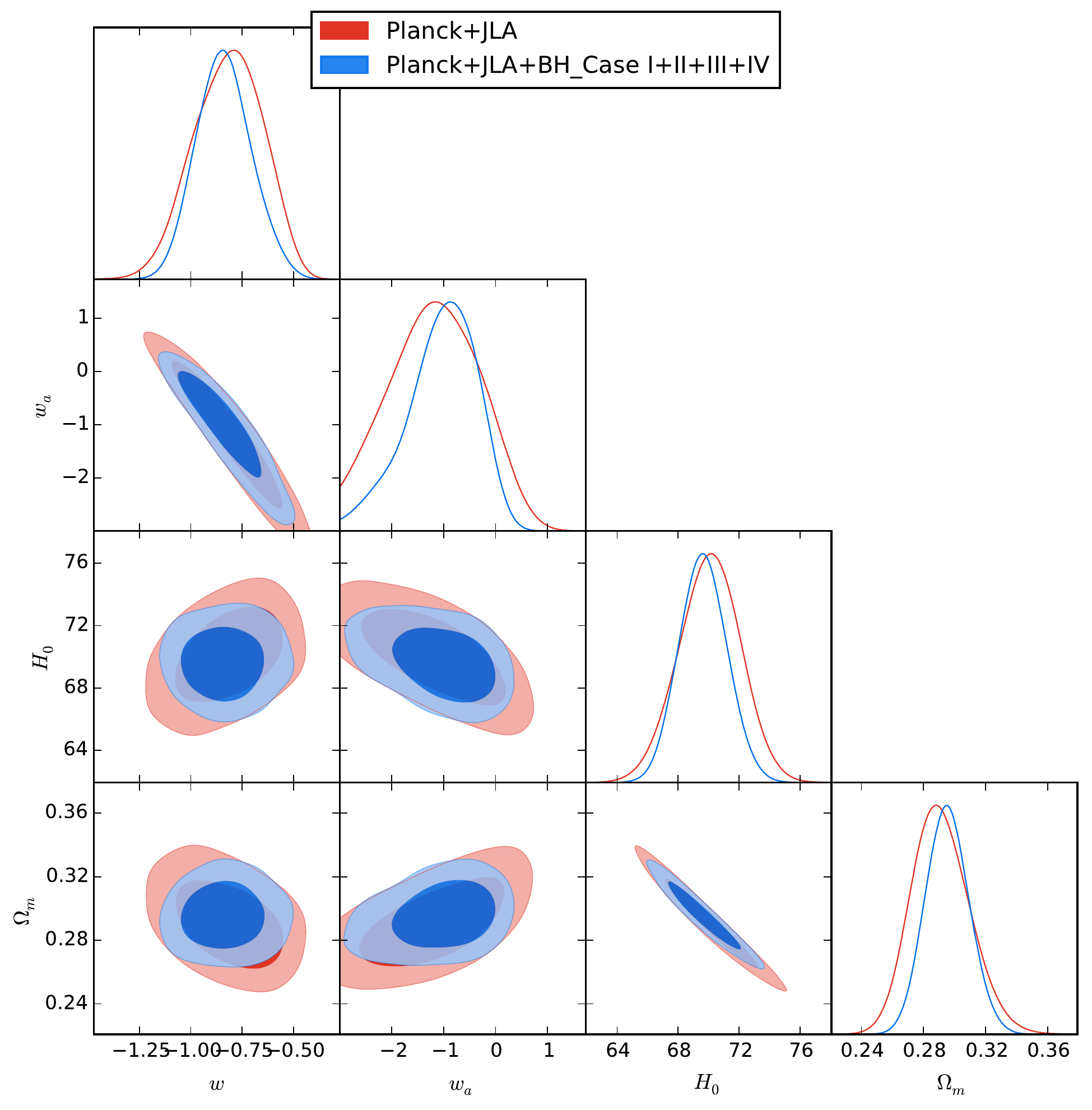}
\includegraphics[width=0.45\textwidth]{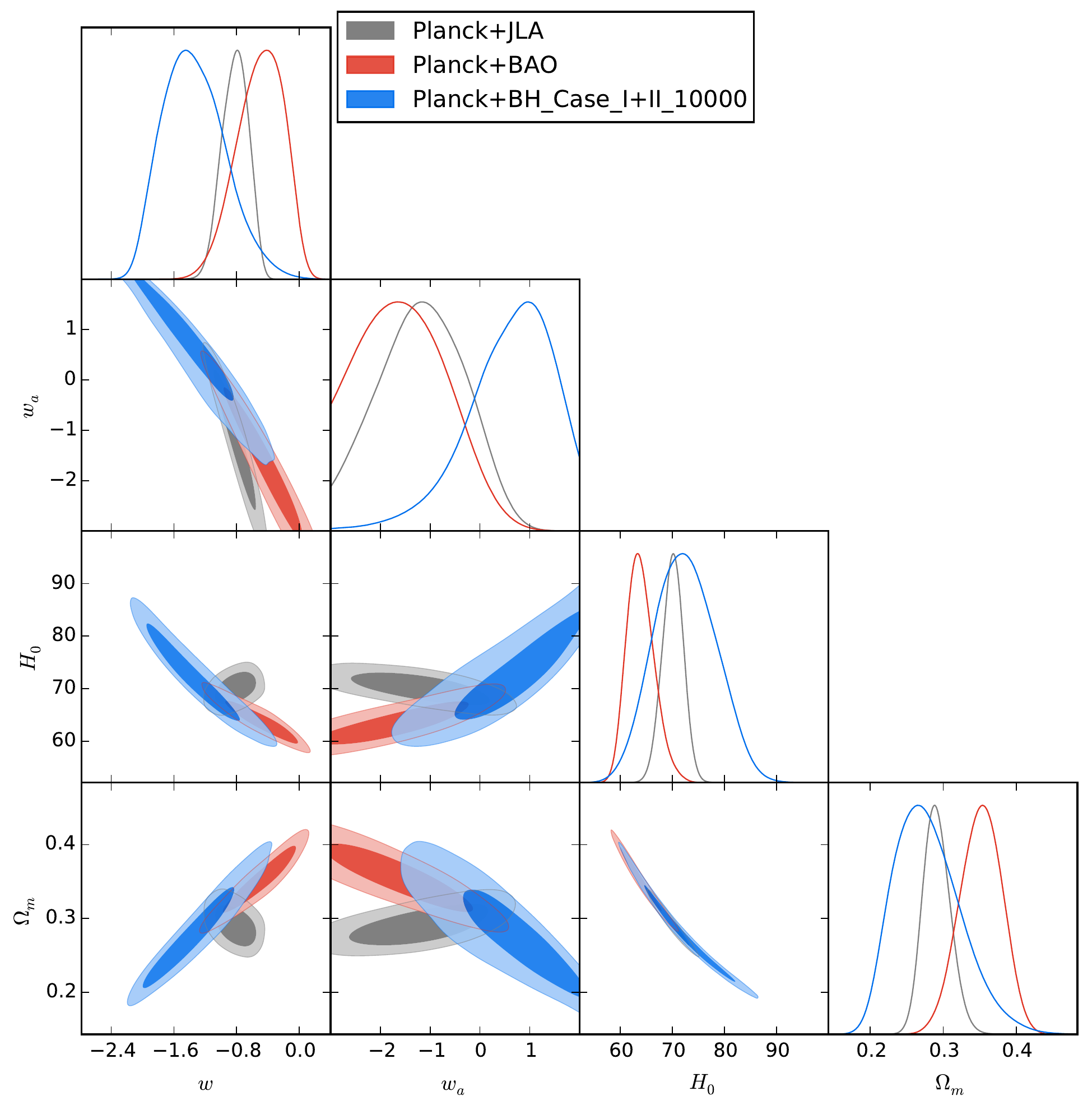}
\caption{The comparison of the constraints of the cosmological parameters from different data sets combinations.}
\label{fig:PJB_IV_all}
\end{figure*}

Obtaining the four cases of the BH mock data sets in the ``\textbf{Part I}'' mock data generating procedure, we use them combined with the {\it Planck} and JLA to constrain the cosmological parameters $H_0$, $\Omega_m$, $w$ and $w_a$ in the CPL parametric dark energy model. With these high redshift BH candles, we expect the tighter constraints of these parameters especially the time-varying equation of state $w_a$. We adopt
Markov chain Monte Carlo (MCMC)~\cite{Lewis:2002ah} method to our analysis. Our parameter constraints are based on the November 2016 version of \textbf{CosmoMC}~\cite{CosmoMC}. Having the four cases of BH candles, we add them to the base data sets {\it Planck} + JLA successively. In cases I, II, and III, we also double the BH candles to see whether the number  of  the BHs has much effect on the constraints of the parameters. We plot all of these parameters' posterior distributions and the counterplots of each combination of two parameters. We also calculate the 68\% C.L. of each parameter and the figure of merit (FOM) for each combination of two parameters to give a quantitative presentation. All of the results are shown in Figs.~\ref{fig:PJB_I} to~\ref{fig:PJB_IV_all} and Tables~\ref{tab:constrain} and~\ref{tab:fom}.

From Figs.~\ref{fig:PJB_I} to~\ref{fig:PJB_IV_all} we can draw some conclusions. First, the BH candles can help to break the degeneracies of the cosmological parameters constrained by  {\it Planck} + JLA as shown in right panel of Fig.~\ref{fig:PJB_IV_all}. Thus the SMAMBHs do have the abilities to improve  the constraints of these cosmological background parameters. Second, the large number of BH candles in the redshift region from 1 to 2 with the 50\% precision are most useful and can make significant contributions to the tighter constraints of the cosmological parameters (see Fig.~\ref{fig:PJB_II}), while the little effect of the higher redshift ($z\sim 2-6$) data sets is due to the fact that dark energy is a small contribution to the energy budget at higher redshift (see Fig.~\ref{fig:PJB_III} and the left panel of Fig.~\ref{fig:PJB_IV_all}). Also, the small number of BH candles with 20\% precision make little contribution to the constraints due to the lack of the data points (see Fig.~\ref{fig:PJB_I}). Finally, from the right panel in Fig.~\ref{fig:PJB_IV_all}, we can see the degeneracy direction of these cosmological parameters for different data sets combinations. Usually, the combination of different data sets can help to break the degeneracy between the cosmological parameters (see the examples for CMB, SNe Ia, and BAO in references~\cite{Suzuki:2011hu,Betoule:2014frx}). Figure~\ref{fig:PJB_IV_all} shows that BAO's result has the same degeneracy direction as the BH candles. Thus both of these two data sets can help to improve the {\it Planck} + JLA's constraints on the background cosmological parameters.

Since the case II of BH candles is the most useful, we calculate the $1-\sigma$ errors and the FOM in the case I + II and case I + II 10000, respectively. The results are shown in Tables~\ref{tab:constrain} and~\ref{tab:fom}, we also include the Planck + JLA for comparison.
We can see that when adding the 10000 BH candles with 50\% precision in redshift $z\sim 1-2$, the $1\sigma$ limits of $w$, $w_a$, $H_0$ and $\Omega_m$ are improved by about 40\%, 60\%, 30\% and 20\%, respectively. For the figure of merit, we can see that the FOMs are almost doubled by these 10000 BH candles except for the $H_0-\Omega_m$. These results show that the high redshift BH candles are more helpful for the constraints of the dynamics of dark energy than $H_0$ and $\Omega_m$.

\begin{table*}[]
\centering

\begin{tabular} {|l|c|c|c|}
\hline
\multicolumn{1}{|c}{Param} &  \multicolumn{3}{|c|}{68\% limits}\\
\hline
\hline
              &\textit{Planck}+JLA
               &\textit{Planck}+JLA+BH Case I+II
               &\textit{Planck}+JLA+BH Case I+II 10000\\
\hline
{\boldmath$w$}
               & $-0.82^{+0.18}_{-0.16}   $
               & $-0.84^{+0.12}_{-0.14}  $
               & $-1.02^{+0.10}_{-0.12}   $ \\

{\boldmath$w_a$}
                 & $-1.15\pm 0.85       $
                 & $-1.01^{+0.76}_{-0.5}     $
                 & $-0.06^{+0.54}_{-0.30}     $ \\

$H_0$
      & $70.1\pm 2.0               $
      & $68.9\pm 1.6               $
      & $67.5\pm 1.4             $   \\

$\Omega_m$
           & $0.291^{+0.017}_{-0.02}            $
           & $0.295\pm 0.014           $
           & $0.316\pm 0.013          $     \\
\hline
\end{tabular}

\caption{\label{tab:constrain} The $1-\sigma$ errors of each parameter for different data set combinations. }
\end{table*}

\begin{table*}[]
\centering

\begin{tabular} {|l|c|c|c|}
\hline
\multicolumn{1}{|c}{Param} &  \multicolumn{3}{|c|}{The FOM of the constraints}\\
\hline
\hline
               &\textit{Planck}+JLA
               &\textit{Planck}+JLA+BH Case I+II
               &\textit{Planck}+JLA+BH Case I+II 10000\\
\hline
$w-w_a$        &20.64
               &26.83
               &38.97\\

$w-H_0$          &3.26
                 &4.83
                 &6.75\\

$w-\Omega_m$ & 354.69
             & 537.10
             & 718.25\\

$w_a-H_0$  & 0.74
           & 1.04
           & 1.52\\

$w_a-\Omega_m$ &77.63
               &111.99
               &157.54\\

$H_0-\Omega_m$ &118.28
               &161.13
               &166.94\\
\hline
\end{tabular}

\caption{\label{tab:fom} The FOM of the constraints for different data set combinations.}
\end{table*}

\section{Project II: Constrain the early expansion rate and dark energy density evolution}
\label{sec:expansion}

Though SEAMBHs candles in redshift $z\sim2-4$ have little improvement on constraining the CPL parametric dark energy parameters, they allow us to constrain the (dimensionless) Hubble parameter $E(z)\equiv H(z)/H_0$ at greater redshifts than previously possible. The Hubble parameter or the expansion history of the Universe is a very important issue in astronomy and cosmology. As indicated in~\cite{Riess:2017lxs}, the quantity $H(z)$ is particularly useful
because it is both a direct probe of cosmology and still
closely tied to the data. As a dynamical quantity, $H(z)$
contains information about the expansion history without reference to any physical cosmological model. Similarly, if we consider the cold dark matter scheme, we can write the evolution of dark energy $\rho_{DE}(z)/\rho_0$ as a function of $E(z)$ from Eq.~\ref{equa:H}, thus we can also constrain the evolution of dark energy density as the Hubble parameter.

\begin{figure*}
\includegraphics[width=0.8\textwidth]{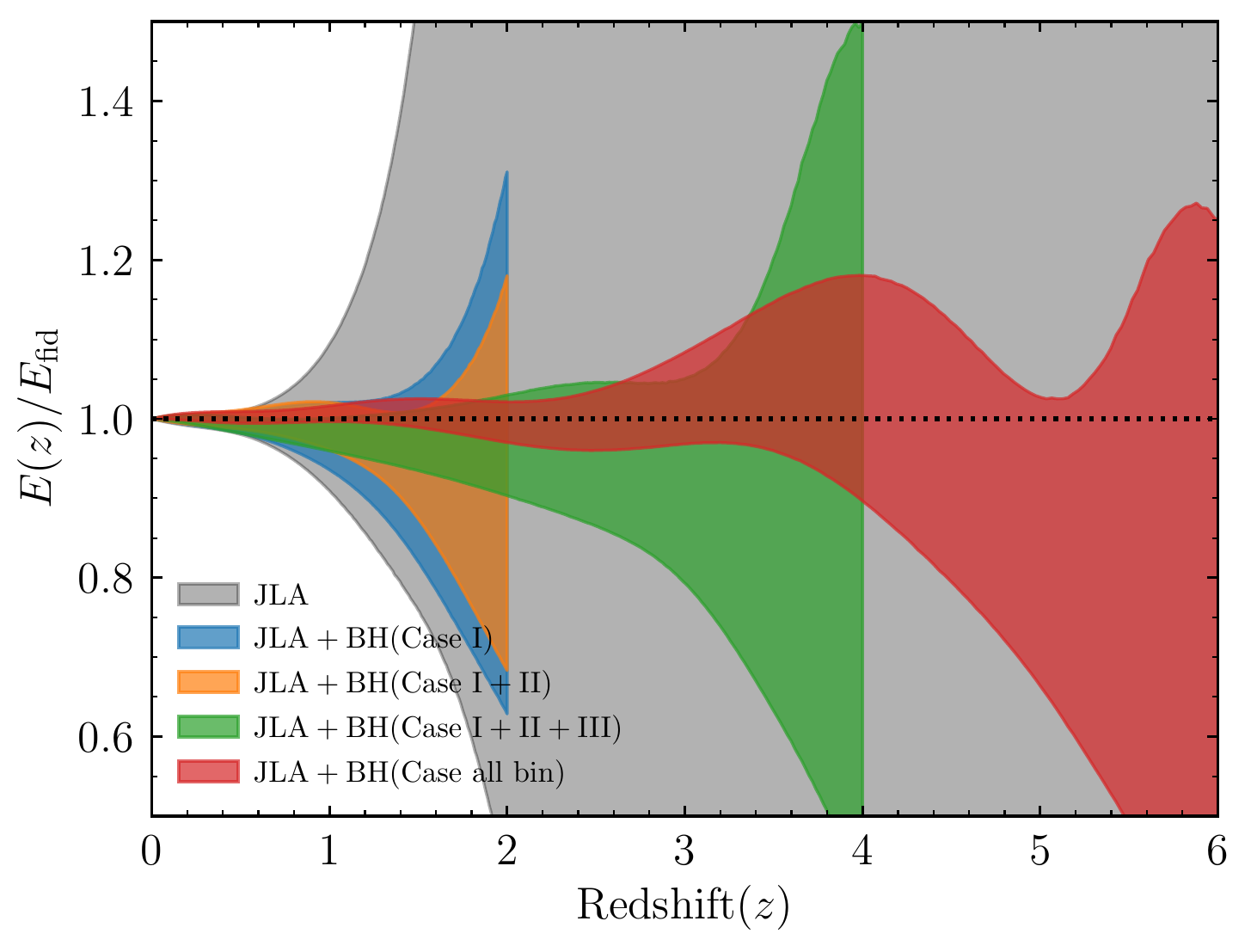}
\caption{The comparisons of the reconstructions of $E(z)/E_{\rm fid}$ from different data set' combinations.}
\label{fig:Ez}
\end{figure*}

In this section, we use the Gaussian process (GP)~\cite{Seikel:2012uu} to smooth the BH candles data sets and reconstruct the $E(z)$ and $\rho_{DE}(z)/\rho_0$. GP is very suitable to apply the distance redshift data to the constraints or reconstructions of parameters in cosmology. Many such works can be found in~\cite{Seikel:2012uu,Cai:2015zoa,Cai:2015pia,Cai:2016vmn}.
Also, we use the JLA as the low redshift source, and add high redshift BH candles in the ``\textbf{Part II}'' procedure as follows: 1) JLA + BH (case I); 2) JLA + BH (case I + II); 3) JLA + BH (case I + II + III); 4) JLA + BH (case all bin). The results in these four steps will indicate the high redshift BH candles' contributions on the constraints of the early expansion rate and the evolution of dark energy, which are shown in Fig.~\ref{fig:Ez}. Note that here we presume a value of $H_0$ since we just want to show the BH candles' abilities on constraining the high redshift expansion rate and dark energy density, and we assume that the value of $H_0$ can be well measured by other observations such as the low redshift SNe Ia.

From Fig.~\ref{fig:Ez} we can see that the adding  BH candles can improve significantly the constraints of Hubble rate in high redshift region. For a comparison, as~\cite{Riess:2017lxs} shows, the uncertainty of $E(z=1.5)$ given by the new high redshift SNe Ia can reduce to $\sim20\%$. While, in the case of BH candles (case I + II + III), the uncertainty can reduce to $\sim 3.9\%$, and even $\sim 1.8\%$ in case all bin, which are much tighter than~\cite{Riess:2017lxs}. We can also find that the constraint of $E(z)$ goes to divergence when the redshift exceeds 1 in the JLA case. But when every case of BH candles is added to the data sets, the redshift of well-constrained $E(z)$ will extend to the higher region ($z\sim 5$ in the case all bin). Every step of adding the BH candles can give tighter constraints on the expansion rate, which indicates that the high redshift SEAMBHs are very powerful for studying the expansion history of the Universe and testing the $\Lambda$CDM model in the high redshift region.

\begin{figure*}
\includegraphics[width=0.8\textwidth]{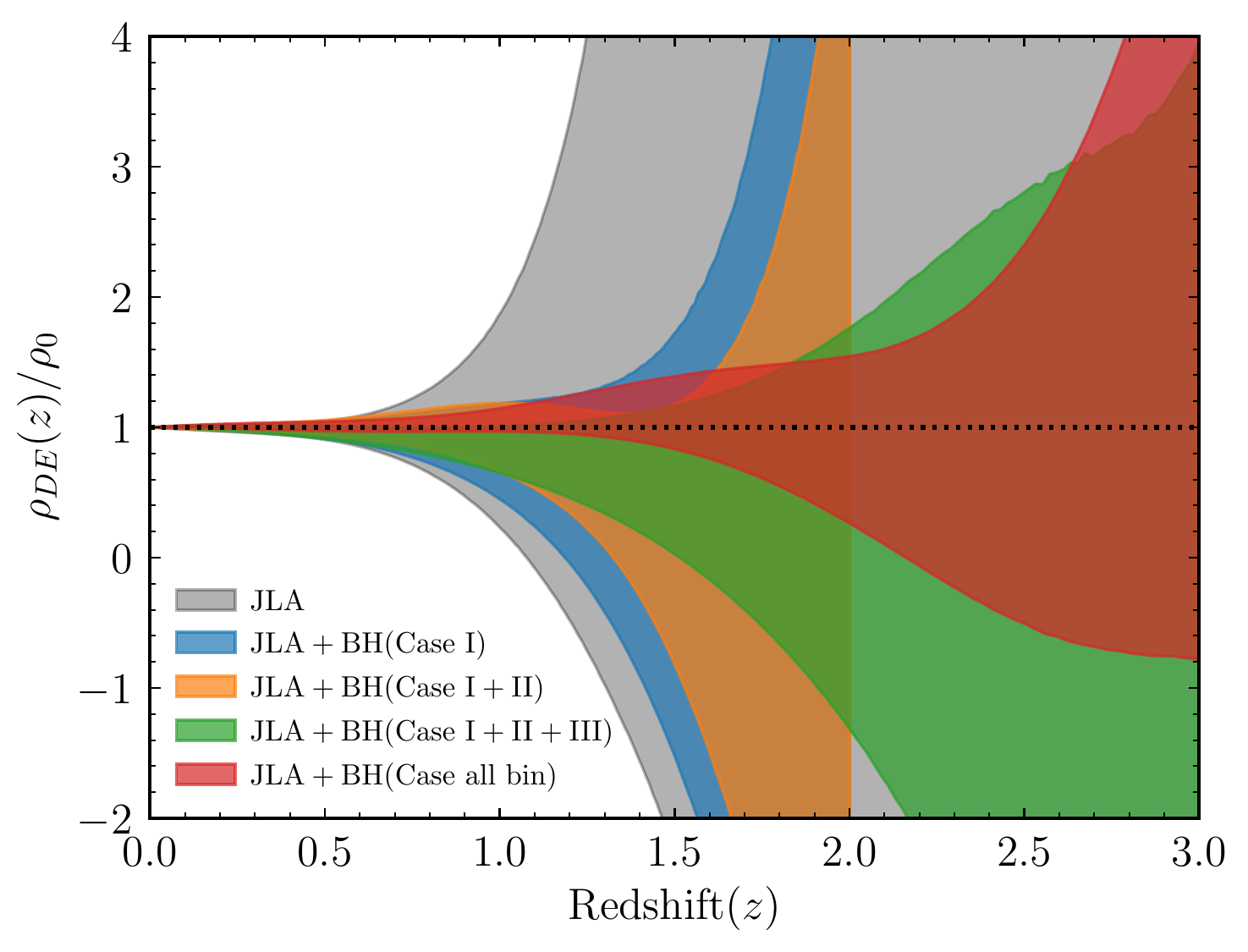}
\caption{The comparisons of the reconstructions of $\rho_{DE}(z)/\rho_0$ from different data set' combinations.}
\label{fig:rhoz}
\end{figure*}

Similarly, the reconstruction of the dark energy density can be also improved much better than with the JLA data. Since the dark energy density is one part of the $E(z)$ in Eq.~\ref{equa:H}, the reconstruction procedure should go further and the quality of the reconstruction is worse than that of $E(z)$. The result is shown in Fig.~\ref{fig:rhoz}. However, we can see significant improvement when the SEAMBHs are added to the data sets.

\section{Project III: Constrain the equation of state $w(z)$ }
\label{sec:P3}

\begin{figure*}
\includegraphics[width=0.8\textwidth]{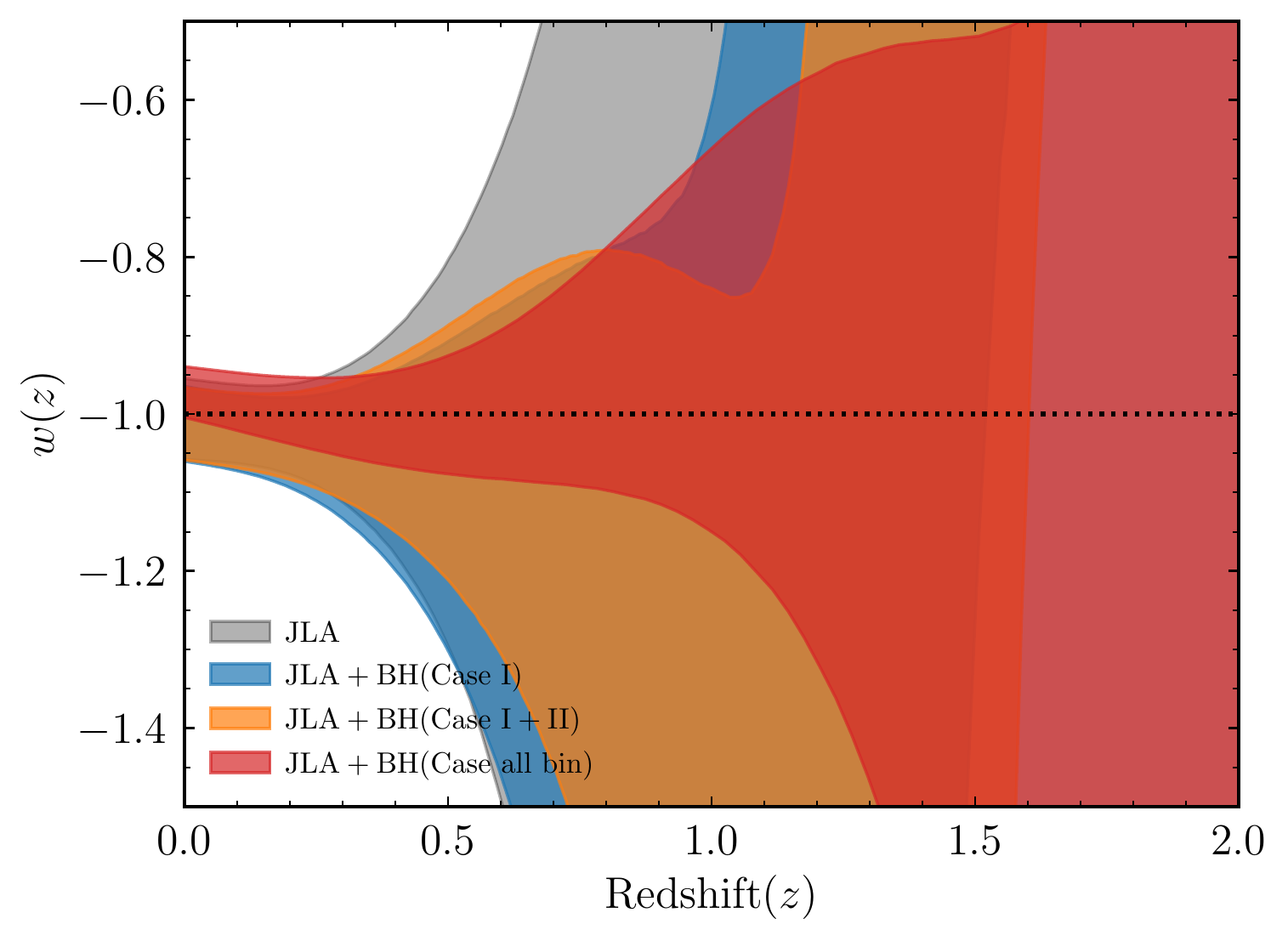}
\caption{The comparisons of the reconstructions of $w(z)$ from different data set combinations.}
\label{fig:wz}
\end{figure*}

In this section, we try to go future again in Eq.~(\ref{equa:H}) and apply the BH candles to a harder issue, that is, we want to reconstruct the equation of state $w(z)$ at any given redshift. We assume the equation of state as a function of $z$, and don't parameterize it using such as the CPL form. As indicated in Eq.~(\ref{equa:w}), $w(z)$ can be determined if the distance redshift relations are given. Here we also assume a flat Universe and the $H_0$ is measured well and fixed. $\Omega_m$ is sampled from the posterior distribution given by {\it Planck} 2015~\cite{Ade:2015xua}.
We focus on whether the high redshift BH candles can give a better reconstruction of  $w(z)$ in higher redshift region. Thus we can measure any evolution of the dynamical dark energy in earlier Universe no matter what the form of the equation of state is. Since the reconstruction of  $w(z)$ is much harder than the Hubble rate and dark energy density, the reconstruction is expected to be worse than that of project II. The results are shown in Fig.~\ref{fig:wz}.

From Fig.~\ref{fig:wz}, we can see that when all of  the BH candles are added (from $z \sim$ 1 to 6), the reconstruction of $w(z)$ can only extend to $z \sim 1.2$ where $w(z)$ is not divergent. Though the full catalogue of BH candles do help to improve the constraints of $w(z)$ compared to the JLA case, it is very hard to obtain a well-reconstructed equation of state at redshift larger than 1. As we can see in Eq.~(\ref{equa:w}),  $w(z)$ is related to the second derivative of the distance, which makes it so hard to reconstruct and constrain the function of equation of state in high redshift region. However, we can see the added BH candles can also improve the reconstruction much better than JLA.

Although it is hard for us to use such BH candles to reconstruct the equation of state in high redshift region, the idea we can use the distance redshift data sets to give directly the evolution of the dynamics of dark energy is notable. If we can measure the distance (mass) of SEAMBHs much more precisely in the future, a much tighter constraint of the equation of state and thus a deep studying of the dynamics of the dark energy at high redshift can be expected.

\section{Conclusions and discussions}
\label{sec:conclusions}

In this work, we simulate the Super-Eddington accreting massive black holes as the candles to probe the cosmology for the first time. The saturated luminosity of SEAMBHs makes them similar to the standard candles like SNe Ia to provide us a new tool for estimating cosmological distance. We simulate the measurements of distance redshift data sets of future SEAMBHs according to the practical estimation of the SEAMBHs distributions from $z\sim 1$ to 6. We use these mock data sets to forecast the abilities of future SEAMBHs candles to probe the cosmology in three different schemes. First, We demonstrate that the SEAMBHs candles can help to break the degeneracies of the cosmological parameters constrained by {\it Planck} and JLA. Thus the SMAMBHs do have the abilities to improve the constraints of cosmological parameters. The large number of BH candles at the redshift from 1 to 2 with the 50\% precision is very useful and with 10000 BH candles the $1\sigma$ limits of $w$, $w_a$, $H_0$ and $\Omega_m$ are improved by about 40\%, 60\%, 30\% and 20\%, respectively. The FOMs of these constraints are doubled except for the $H_0-\Omega_m$ which indicates these high redshift SEAMBHs candles are more helpful to the constraints of equation of state. Second, adding the SEAMBHs data to JLA, we can significantly extend the well constraints of expansion rate of the Universe and the evolution of dark energy density to much higher redshift. This shows the powerful potential of SEAMBHs candles on studying the expansion history of the Universe and test of $\Lambda$CDM model at high redshift. Finally, we also try to use these SEAMBHs candles to reconstruct the equation of state. Our results show that it is very hard to extend the redshift of well-reconstructed $w(z)$ to larger than 1. This is due to the fact that the reconstruction of $w(z)$ is determined by the high order derivatives of the distance. Nevertheless, with the more precise measurements of SEAMBHs' distances in the future, this reconstruction to detect any form of an evolving equation of state of dark energy can be possible in the high redshift region.

In summary, SEAMBHs can serve us as a new and independent source to probe the cosmology. For the large number of sources in high redshift region even to $z\sim 7$, SEAMBHs can play very important roles in the studying of dynamical dark energy, early expansion history of Universe and tests of the cosmological model at high redshift.

\begin{acknowledgments}
We would like to thank Jian-Min Wang and Pu Du for their helpful discussions and advices
on these simulations.
RGC is supported by the National Natural Science Foundation of China Grants No.11690022, No.11435006 and No.11647601, and by the
Strategic Priority Research Program of CAS Grant No.XDB23030100 and by the Key Research Program of Frontier Sciences of CAS.
ZKG is supported by the National Natural Science Foundation of China Grants No.11690021, No.11575272 and No.11335012.
QGH is supported by grants from NSFC (grant NO. 11335012, 11575271, 11690021, 11647601), Top-Notch Young Talents Program of China, and partly supported by Key Research Program of Frontier Sciences, CAS.
TY is supported by the National Natural Science Foundation of China Grants No. 210100088 and No.  210100086, and by China Postdoctoral Science Foundation under grant No. 2017M620662.

\end{acknowledgments}

\end{document}